\documentclass[journal]{ASD}
\renewcommand\figurename{Figure}
\renewcommand\refname{References}
\renewcommand\tablename{Table}
\IEEEoverridecommandlockouts

\usepackage[utf8]{inputenc} 
\usepackage[T1]{fontenc}

\usepackage{cite}

\ifCLASSINFOpdf
  \usepackage[pdftex]{graphicx}
\else
\fi
\usepackage[cmex10]{amsmath}
\usepackage{multirow}
\usepackage{array}
\usepackage[lofdepth,lotdepth]{subfig}

\begin{document}


%
\title{Performance Comparison of Balanced and Unbalanced Cancer Datasets using Pre-Trained Convolutional Neural Network}

\author{\IEEEauthorblockN{Ali Narin$^{1}$}\\
\IEEEauthorblockA{$^{1}$Department of Electrical and Electronics Engineering, Zonguldak Bulent Ecevit University, Zonguldak, Turkey\\}
{alinarin}@beun.edu.tr}

%

\maketitle

\begin{abstract}
Cancer disease is one of the leading causes of death all over the world. Breast cancer, which is a common cancer disease especially in women, is quite common. The most important tool used for early detection of this cancer type, which requires a long process to establish a definitive diagnosis, is histopathological images taken by biopsy. These obtained images are examined by pathologists and a definitive diagnosis is made. It is quite common to detect this process with the help of a computer. Detection of benign or malignant tumors, especially by using data with different magnification rates, takes place in the literature. In this study, two different balanced and unbalanced study groups have been formed by using the histopathological data in the BreakHis data set. We have examined how the performances of balanced and unbalanced data sets change in detecting tumor type. In conclusion, in the study performed using the InceptionV3 convolution neural network model, 93.55\% accuracy, 99.19\% recall and 87.10\% specificity values have been obtained for balanced data, while 89.75\% accuracy, 82.89\% recall and 91.51\% specificity values have been obtained for unbalanced data. According to the results obtained in two different studies, the balance of the data increases the overall performance as well as the detection performance of both benign and malignant tumors. It can be said that the model trained with the help of data sets created in a balanced way will give pathology specialists higher and accurate results.
\end{abstract}



%
\IEEEpeerreviewmaketitle

\IEEEpubidadjcol

\section{INTRODUCTION}

Cancer is a very common disease today that is caused by the uncontrolled proliferation of cells \cite{1}. Cancer, which is often fatal, has become a serious public health problem. One of the most common types of cancer is breast cancer seen in men and women \cite{2}. This cancer, which is frequently seen in breast cells in women, causes hardness and lump in the breast. These structures need to be examined by various imaging methods and microscopically. As a result of the examination, it can be definitively determined whether the structure formed is benign or malignant \cite{3}.
Early detection of the cancerous structure is very important. Because the spread of cancer cells to other cells and tissues can be prevented by early detection. Pathologists who examine tissue samples at different magnifications can make the definitive diagnosis. Performing this tiring and demanding process with computer-aided diagnosis (CAD) systems will reduce the workload of pathologists and accelerate early diagnosis \cite{4}. Thus, a person with cancer can be intervened earlier.
In the literature, detection of breast cancer type with CAD systems is quite common. Besides traditional methods, deep learning models are widely used. Among traditional methods, it is seen that k-nearest neighbor, Naive Bayes, decision trees, back propagation multi-layer perceptron networks, probabilistic neural networks and support vector machine algorithms are mostly used as classifiers \cite{5,6,7,8}.
In addition to various state-of-art architectures among deep learning-based models, studies have also been carried out with special architectures \cite{9,10,11,12}.

In this study, a performance comparison was made with a balanced and unbalanced data set obtained from benign / malignant histopathological breast cancer images. The study was carried out end-to-end using the pre-trained InceptionV3 model. In the study, external pre-processing and manual feature extraction were not performed. According to the results obtained, it has been observed that the balanced data set has a higher performance.

In the rest of the article, dataset and its features, the pre-trained convolutional neural network model architecture, performance metrics and the experimental results obtained will be included. Finally, discussion of the findings will be given.

\section{MATERIAL AND METHODS}

\subsection{Dataset}

In this study, breast cancer images were obtained from the "Breast Cancer Histopathological Image Classification (BreakHis)" (https://web.inf.ufpr.br/vri/databases/breast-cancer-histopathological-database-breakhis/) dataset that is accessible to everyone \cite{9}. This dataset contains 7909 breast cancer histopathological images from 82 patients. While 2480 of 7909 images belong to the benign class, 5429 of them belong to the malignant class. The distributions of the data with 4 different magnifications (40X, 100X, 200X and 400X) are given in Table 1. In addition, each data found in the data set is 700x460 in size and 8 bit deep. Image examples of different magnification amounts in the data set are shown in Figure 1.

\begin{figure}[htbp]
\centering
\includegraphics[width=0.75\columnwidth]{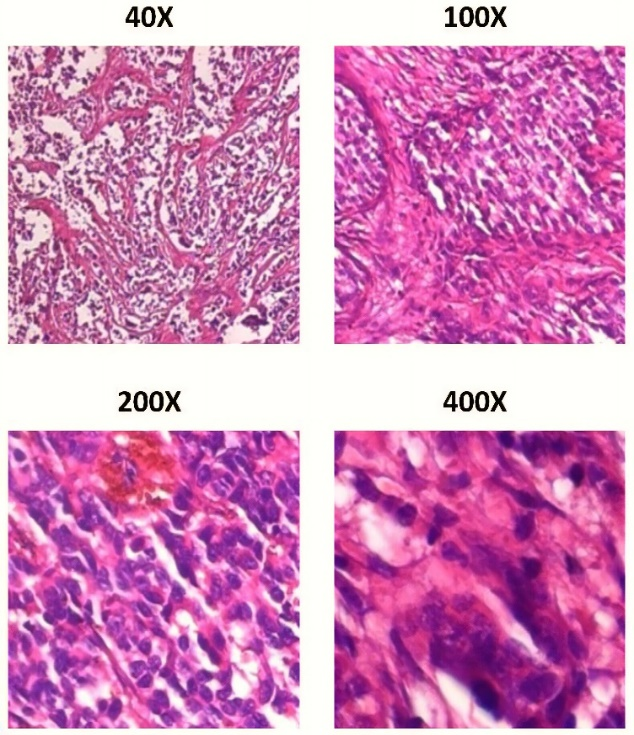}
\caption{Examples of histopathological images of different magnifications.}
\label{figure1}
\end{figure}

\begin{table}[htbp]
  \centering
  \begin{tabular}{|c|c|c|c|}
\hline
Magnification & Benign & Malignant & Total\\
\hline
40X		&625		&1370	&1995\\
100X	&644		&1437	&2081\\
200X	&623		&1390	&2013\\
400X	&588		&1232	&1820\\
Total	&2480	    &5429	&7909\\
\hline
\end{tabular}
\caption{Details about the whole dataset.}
\label{tablo}
\end{table}

\subsection{Pre-Trained Convolutional Neural Networks}

Convolutional neural networks (CNN) are a type of deep architectures of artificial neural networks. It mostly performs high on image data. It was proposed by inspiring from the eyesight of animals. The most important aspect of these models is that they provide direct feature extraction on images. Thus, there is no need for an external feature extraction process. Generally, a convolution neural network consists of convolution layers, pooling layers, and fully connected layers. Dropout and/or normalization layers can also be added to protect the model from over-learning problem and to increase performance. Activation functions are used to transfer image properties to the network.

A lot of data is needed to train convolutional neural network models. In cases where the number of data is not sufficient, pre-trained models are preferred. Also, pre-trained models are used because training a model from scratch is very time consuming. In this study, pre-trained InceptionV3 CNN model is used. This model uses inception modules and aims to experiment with different convolution combinations to increase its performance by diversifying its properties. The Inception network is an architecture defined as the network within the network. It is based on the simultaneous performance of filtering and pooling processes in the convolution layers. It performs operations in modules \cite{13}.

This study was carried out by adding 3 layers to the pre-trained InceptionV3 model as shown in Figure 2.

\begin{figure}[htbp]
\centering
\includegraphics[width=0.95\columnwidth]{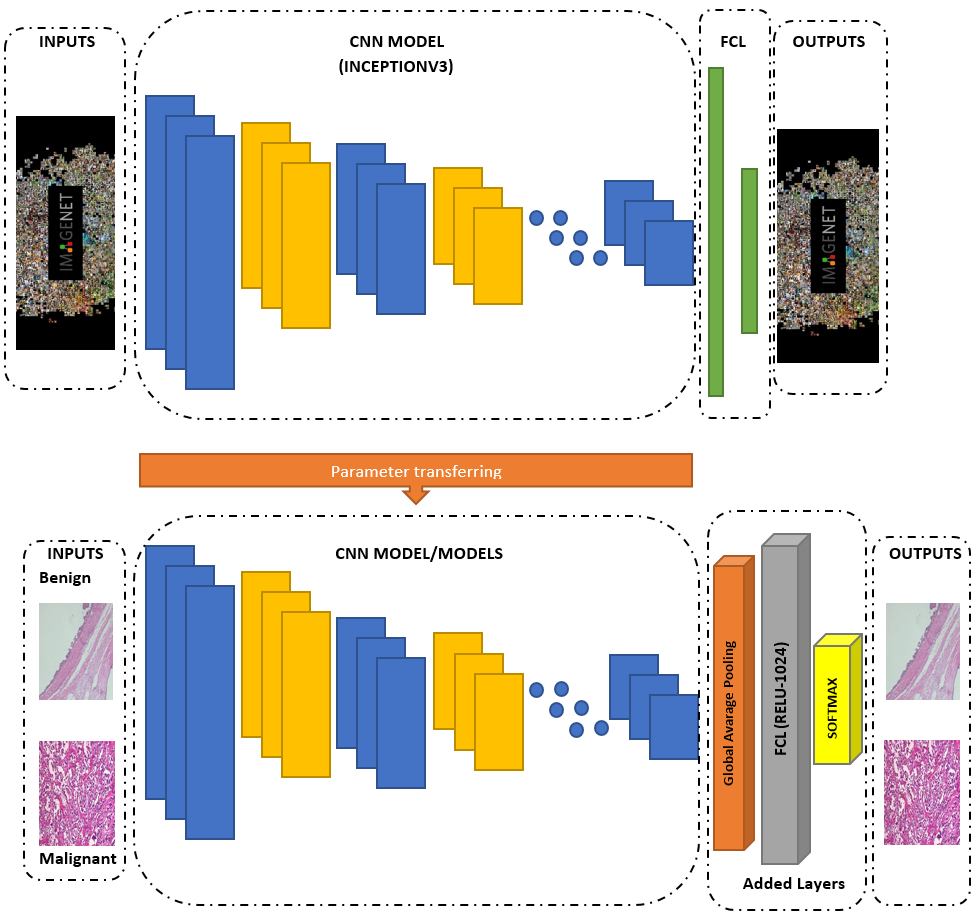}
\caption{Demonstration of Pre-Trained InceptionV3 model.}
\label{figure2}
\end{figure}

\subsection{Metrics}

Three metrics were used for the performance of the deep learning model [14][18]. 

\begin{eqnarray}
  	Recall(REC)&=&\frac{TP}{TP+FN}\\
	Specificity(SPE)&=&\frac{TP}{TP+FP}\\
	Accuracy(ACC)&=&\frac{TP+TN}{TP+TN+FP+FN}
\end{eqnarray}

Here, TP, True positive; TN, true negative; FN, false negative and FP, false positive are called. If these values are taken into consideration for our problem, the detection of a benign (positive) tumor as benign (TP) and the detection of a malignant (negative) tumor as malignant (TN) are indicated. Detection of a benign tumor as malignant (FP) and detection of a malignant tumor as benign (FN) are expressed \cite{17,18}.




\section{RESULTS}

In this study, Python programming language was used to establish the model and get results. The entire study was carried out using the Tensorflow-Keras library on Google Colab \cite{15}. The BreakHis data set is arranged as balanced and unbalanced as in Table 2 and Table 3.

\begin{table}[htbp]
  \centering
  \begin{tabular}{|c|c|c|c|}
\hline
Magnification & Benign & Malignant & Total\\
\hline
40X		&625		&1370	&1995\\
100X	&644		&1437	&2081\\
200X	&623		&1390	&2013\\
400X	&588		&1232	&1820\\
Total	&2480	    &5429	&7909\\
\hline
\end{tabular}
\caption{Details of the unbalanced dataset.}
\label{tablo}
\end{table}

\begin{table}[htbp]
  \centering
  \begin{tabular}{|c|c|c|c|}
\hline
Magnification & Benign & Malignant & Total\\
\hline
40X		&625		&625	&1250\\
100X	&644		&644	&1288\\
200X	&623		&623	&1246\\
400X	&588		&588	&1176\\
Total	&2480	    &2480	&4960\\
\hline
\end{tabular}
\caption{Details of the balanced dataset.}
\label{tablo}
\end{table}

Here, the class with the least number of data is based on the balanced and unbalanced arrangement of the data. A balanced data set was formed with randomly selected data from the data with a large number of data. It has not been studied separately according to the magnification amounts. The results of the study were found by gathering data belonging to the same class in a single data set. In the unbalanced data set, there are data of 5429 malignant tumors against 2480 data of benign tumors. Balanced data set includes 2480 malignant tumor data against 2480 benign data.

90\% of the data was used to train the model, while 10\% of the data was used for testing. Training was carried out for 30 epochs to protect model training from over-fitting. Optimization of the weights in the training of the models was done by Adaptive Moment Estimation (ADAM) algorithm. Learning rate parameter (LR) and batch size (BH) parameters were determined as 0.00003 and 30, respectively.
The results obtained over two different data sets are given in Table 4.

\begin{table}[htbp]
  \centering
  \begin{tabular}{|c|c|c|c|c|c|c|c|}
\hline
Dataset	&TP	&TN	&FP	&FN	&ACC&REC&SPE\\
Type& & & & &(\%)&(\%)&(\%)\\
\hline
Unbalanced	&213	&496	&46	&35	&89.75	&85.89	&91.51\\
\hline
Balanced	&246	&218	&30	&2	&93.55	&99.19	&87.90\\
\hline
\end{tabular}
\caption{Comparison of balanced and unbalanced data set performance.}
\label{tablo}
\end{table}

Considering the results given in Table 4, It is seen that malignant tumors, which are more numerous in the unbalanced data set, increase the specificity value, while the recall value is negatively affected and its performance decreases. In the balanced data set, it is seen that the overall performance is 93.55\% and the recall value is quite high. In general, it can be said that the number of data in the data set and the number of data belonging to the classes have a significant effect on the performances.

\section{DISCUSSION}
Studies conducted on the histopathological images dataset of BreakHis breast cancer are given in Table 5. Many of these recent studies use deep learning models. In these studies, magnification factor based performances are given.

Deniz et al. obtained the highest accuracy of 91.37\% with 200X magnification data. Spanhol et al. stated that reached 85.6\% accuracy values at 40X magnification for AlexNet, and the highest 90.3\% performance value at 200X magnification in another study. Bayramoğlu et al. found an accuracy of 84.63\% at 200X magnification. Gour et al. obtained an accuracy of 92.52\% at 200X magnification. [9, 10, 11, 12, 16].

In this study, the highest general accuracy value of 93.55\% was achieved with a balanced data set. In this respect, higher performance has been obtained than many studies in the literature. In addition, collecting the same class of data with all magnifications and performing the study is among the prominent aspects of the article.
It is predicted that the achievements will increase more by increasing the data set used in the study. In future studies, working with different CNN models and parameters can be carried out in more detail.

\begin{table}[htbp]
  \centering
  \begin{tabular}{|c|c|c|c|}
\hline
Author(s) &Magnification &Methods &ACC(\%)\\
\hline
\cite{16}  	&200X	&AlexNet and VGG16	&91.37\\
\cite{9}  	&40X	&AlexNet				&85.60\\
\cite{10}  	&200X	&deCAF + CaffeNet	&90.30\\
\cite{11}  	&200X	&Multi-Task CNN		&84.63\\
\cite{12}  	&100X	&ResHist				&84.34\\
		&200X	&Data Aug. + ResHist	&92.52\\
This Study	&All (Unbalanced)	&Pre-trained InceptionV3	&89.75\\
This Study	&All (Balanced)	&Pre-trained InceptionV3		&93.55\\
\hline
\end{tabular}
\caption{Comparison of the obtained results with other studies in the literature conducted with the same data set.}
\label{tablo}
\end{table}

\end{document}